\documentclass[conference]{IEEEtran}
\IEEEoverridecommandlockouts
\usepackage{multirow}
\usepackage{cite}
\usepackage{amsmath,amssymb,amsfonts}
\usepackage{algorithmic}
\usepackage{graphicx}
\usepackage{textcomp}
\usepackage{xcolor}
\def\BibTeX{{\rm B\kern-.05em{\sc i\kern-.025em b}\kern-.08em
    T\kern-.1667em\lower.7ex\hbox{E}\kern-.125emX}}
\begin{document}

\title{Towards a Network Expansion Approach for Reliable Brain-Computer Interface\\

\thanks{This work was supported by the National Research Foundation of Korea (NRF) grant funded by the MSIT (No. 2022-2-00975, MetaSkin: Developing Next-generation Neurohaptic Interface Technology that enables Communication and Control in Metaverse by Skin Touch) and the Institute of Information \& Communications Technology Planning \& Evaluation (IITP) grant, funded by the Ministry of Science and ICT (MSIT) (RS-2019-II190079, Artificial Intelligence Graduate School Program).}}

\author{\IEEEauthorblockN{Byeong-Hoo Lee }
\IEEEauthorblockA{\textit{Dept. of Brain and Cognitive Engineering} \\
\textit{Korea University}\\
Seoul Republic of Korea \\
bh$\_$lee@korea.ac.kr}
\and
\IEEEauthorblockN{Kang Yin}
\IEEEauthorblockA{\textit{Dept. of Artificial Intelligence} \\
\textit{Korea University}\\
Seoul Republic of Korea \\
charles$\_$kang@korea.ac.kr}
}
\maketitle

\begin{abstract}
Robotic arms are increasingly being used in collaborative environments, requiring an accurate understanding of human intentions to ensure both effectiveness and safety. Electroencephalogram (EEG) signals, which measure brain activity, provide a direct means of communication between humans and robotic systems. However, the inherent variability and instability of EEG signals, along with their diverse distribution, pose significant challenges in data collection and ultimately affect the reliability of EEG-based applications. This study presents an extensible network designed to improve its ability to extract essential features from EEG signals. This strategy focuses on improving performance by increasing network capacity through expansion when learning performance is insufficient. Evaluations were conducted in a pseudo-online format. Results showed that the proposed method outperformed control groups over three sessions and yielded competitive performance, confirming the ability of the network to be calibrated and personalized with data from new sessions.

\end{abstract}


\section{Introduction}
Recent deep learning-based artificial intelligence has demonstrated human-level cognitive abilities \cite{lee2020uncertainty, badrinarayanan2017segnet, lee2018deep}. They have enabled robotic arms to physically engage with humans. To operate effectively and safely, these robotic arms must accurately interpret human control intentions. A promising approach to achieve this is the decoding of brain signals, specifically electroencephalogram (EEG) signals, which can be acquired non-invasively \cite{ravi2022enhanced, penaloza2018bmi}. This makes EEG a viable option for controlling external devices \cite{lee2020continuous, mao2019brain, han2020classification, hsu2015real}. The brain-computer interface (BCI) framework uses EEG signals as control inputs, and this study explores its application to robotic arm manipulation. A widely used EEG elicitation paradigm is motor imagery (MI) \cite{mcfarland1997spatial}. MI involves imagining muscle movements without physical action.

Despite the promise of BCI technology, several challenges remain: (1) The inherent complexity and variability of EEG signals require frequent calibration of decoding algorithms \cite{prabhakar2020framework}. This variability can lead to wide data distributions, complicating both calibration for new session datasets ultimately affecting performance. (2) User-related factors, such as fatigue and concentration, further contribute to inconsistencies in EEG signal generation \cite{monteiro2019using, jeong2019classification, yang2017mind}, making it difficult to extract fundamental features. (3) The results of (1) and (2) indicate that shallow convolutional neural networks (CNNs) struggle to achieve sufficient performance in classifying MI related to similar body movements, such as hand gestures.

To address these challenges, we propose an extensible decoding algorithm tailored for EEG-driven robotic arm control. In each session, users wear an EEG cap, control the robotic arm via EEG signals, and then remove the cap. Our goal is to improve classification performance, especially for new sessions within each session. As data accumulates, we also explore the potential for personalizing the decoding algorithm \cite{kim2018discriminative}. A pipeline is introduced that expands the capacity of the neural network through strategic network expansion. To mitigate indiscriminate expansion, we employ selective initialization and training. Our baseline evaluations use publicly available datasets, followed by 3 sessions with novice users for data collection.

The following constitutes a list of the major contributions of this research: (1) By introducing a network expansion strategy to shallow CNNs, we established a pipeline that allows the model to adapt to the variability of EEG signals. (2) We conducted evaluations over three sessions to assess the generalizability of the network expansion strategy and visualized the extracted features to explore its impact on feature extraction. (3) The experimental results confirm that our proposed method is a cost-effective solution for practical EEG-based robotic arm control using advanced machine learning techniques.

\section{Related Works}
In recent years, a number of research efforts have significantly advanced the classification of endogenous BCIs. For example, Lawhern \textit{et al.} \cite{eegnet} implemented deep and separable convolutions to effectively summarize individual features over time, using more channel information in the process. Their results showed that CNNs could be trained with a reduced number of parameters when applied to BCIs. Sakhavi \textit{et al.} \cite{sakhavi} proposed an innovative temporal representation approach that combined FBCSP with CNN architectures to gain more temporal insights, recognizing the high temporal resolution inherent in EEG signals. In addition, Fahimi \textit{et al.} \cite{Fahimi} introduced a deep convolutional generative adversarial network framework to generate artificial EEG signals, demonstrating that data augmentation through generative models can improve classification performance. Collectively, these studies have led to remarkable improvements in classification tasks.

Several studies have explored multi-session and multi-paradigm methods in BCIs. For example, in addition, Thomas \textit{et al.} \cite{thomas2013analysis} focused on evaluation metrics for complex MI BCIs that incorporate elements such as adaptive classification, error detection and correction, signal fusion, and shared control. Their work used both simulated and experimental data, while also reviewing existing literature to better understand BCI evaluation, particularly the relationship between data usage and the specific BCI components being analyzed. Lee \textit{et al.} \cite{lee2019eeg} presented a data set tailored for BCI systems that includes three primary paradigms: MI, event-related potentials, and steady-state visually evoked potentials. This dataset was collected from a diverse group of subjects over multiple sessions and includes both psychological and physiological variables of the users. The researchers evaluated decoding accuracies for each paradigm and analyzed performance variations across subjects and sessions.

\begin{figure}[!t]
  \centerline{\includegraphics[scale=0.95]{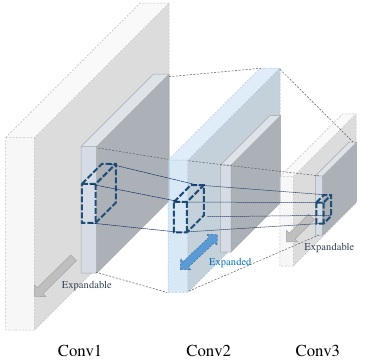}}
  \caption{Layer expansion visualization. This example illustrates the expansion of the middle layer among the three convolutional layers, resulting in an increase in the number of convolutional channels as well as an enlargement of the feature size.}
  \label{fig1}
\end{figure}

\section{Methods}
In this research, we examine the capability of a neural network to improve classification performance. For our investigation, we utilized the hand task MI dataset \cite{jeong2020multimodal}, which encompasses classes categorized as ``cylindrical,'' ``spherical,'' and ``lumbrical.'' Each data sample is represented in the form $ x \in \mathbb{R}^{C \times T} $, with the data being downsampled to a frequency of 250 Hz, resulting in a total of 1000 time points per sample. Training and evaluation conducted subject-independent manner to ensure sufficient training dataset.

In order to accommodate the variations in EEG patterns, a three-layer CNN was designed with the capability to expand its architecture dynamically. The preliminary stage of our training process entails a sparse training methodology, wherein the network is trained with constrained complexity to guarantee effective learning. Subsequently, we incrementally augment the network's capacity to prevent superfluous and excessive expansion. To this end, we integrate an elementwise $L_1$ norm regularization into the loss function. This regularization technique imposes a penalty on the weights of the network, encouraging sparsity and ultimately leading to a more efficient model. The loss function can be mathematically expressed as follows:

\begin{equation}
\mathcal{L} = -\frac{1}{N} \sum_{i=1}^{N} \hat{y} \log(f(x)) + \lambda \sum_{j=1}^{J} \vert W^j_{\text{initial}} \vert,
\end{equation}
where $ W^j $ represents the weights associated with the $ j^{th} $ layer of the network, $ \lambda $ denotes the regularization parameter, and $ J $ is the total number of layers in the neural architecture.

\begin{table}[]
\centering
\renewcommand{\arraystretch}{1}
{
\caption{Design choices for the experiments. ELU denotes elu activation function. The convolutional layers are expandable.}
\resizebox{\columnwidth}{!}{%
\begin{tabular}{c|cccc}
\hline
Network                    & Layer \#            & Type                         & Parameter                 & Output size                \\ \hline
\multirow{20}{*}{$f_{cs}$} & 1                  & Input                        & -                       & 58$\times$1000                    \\ \cline{2-5} 
                           & \multirow{3}{*}{2} & \multirow{3}{*}{Convolutional} & \# of channels={[}1,56{]}    & \multirow{3}{*}{56$\times$58$\times$970} \\
                           &                    &                              & Kernel size={[}1,32{]}    &                            \\
                           &                    &                              & Batch norm = 56           &                            \\ \cline{2-5} 
                           & \multirow{4}{*}{3} & \multirow{4}{*}{Convolutional} & \# of channels={[}56,112{]}  & \multirow{4}{*}{112$\times$1$\times$970} \\
                           &                    &                              & Kernel size={[}58,1{]}    &                            \\
                           &                    &                              & Batch norm = 112          &                            \\
                           &                    &                              & ELU                       &                            \\ \cline{2-5} 
                           & \multirow{3}{*}{4} & \multirow{3}{*}{Pooling}     & AvgPool2d = {[}1,2{]}     & \multirow{3}{*}{112$\times$1$\times$485} \\
                           &                    &                              & Stride = {[}1,2{]}        &                            \\
                           &                    &                              & Dropout (p=0.5)           &                            \\ \cline{2-5} 
                           & \multirow{4}{*}{5} & \multirow{4}{*}{Convolutional} & \# of channels={[}112,224{]} & \multirow{4}{*}{224$\times$1$\times$454} \\
                           &                    &                              & Kernel size={[}1,32{]}    &                            \\
                           &                    &                              & Batch norm =224           &                            \\
                           &                    &                              & ELU                       &                            \\ \cline{2-5} 
                           & \multirow{3}{*}{6} & \multirow{3}{*}{Pooling}     & AvgPool2d = {[}1,2{]}     & \multirow{3}{*}{224$\times$1$\times$227} \\
                           &                    &                              & Stride = {[}1,2{]}        &                            \\ 
                           &                    &                              & Dropout (p=0.5)           &                            \\ \cline{2-5} 
                           & 7                  & Linear                       & Linear (*, 224)           & 224$\times$1$\times$224                  \\ \cline{2-5} 
                           & 8                  & FC layer                     &           -                & 50176                      \\ \cline{2-5} 
                           & 9                  & Classification                     &  Softmax (50176, 6)                         & 1$\times$6                      \\ \hline
\end{tabular}}}
\end{table}

\begin{figure*}[!t]
  \centerline{\includegraphics[scale=0.83]{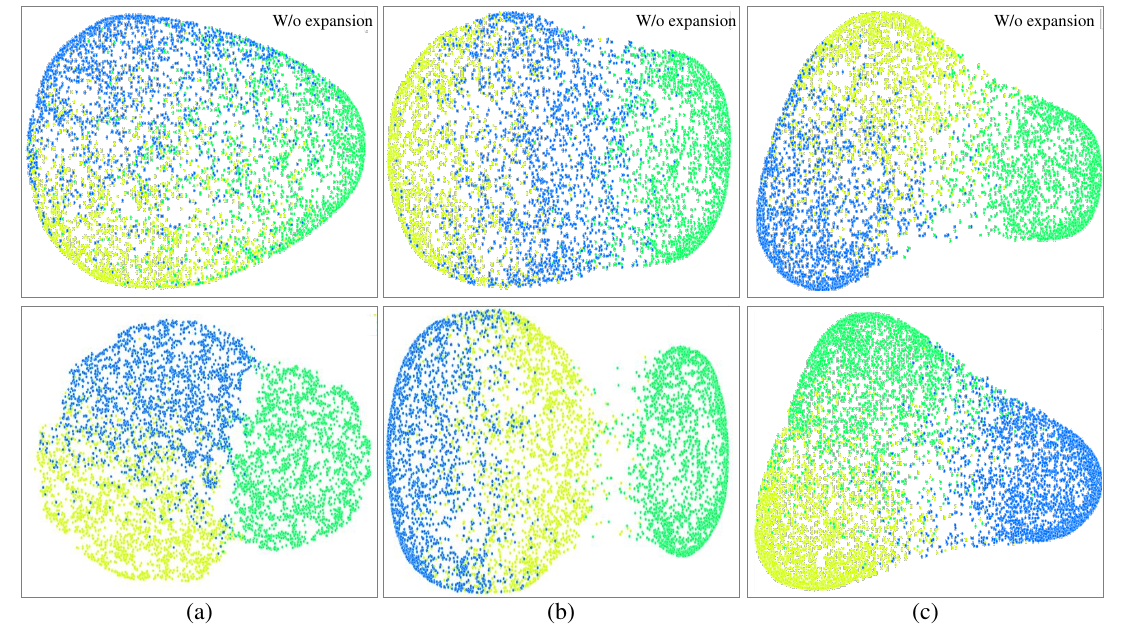}}
  \caption{The visualization of feature extraction is demonstrated using t-SNE \cite{van2008visualizing}. The top row illustrates  the results for the datasets of (a) Session 1, (b) Session 2, and (c) Session 3 without network expansion, while the bottom row displays the visualization of the features with network expansion applied.}
  \label{fig1}
\end{figure*}

In the event that the computed loss value exceeds a predetermined threshold during the training process, this indicates that the capacity of the network in question requires augmentation. In response to this indication, additional nodes are introduced into the existing layers, thereby expanding the network weights as follows: The weight matrix is represented as $W = [W; \hat{W}]$, where $W$ represents the existing weights and $ \hat{W} $ denotes the weights associated with the newly added nodes. Given the flexibility to integrate an arbitrary number of $o$ nodes into the network, the objective function must account for optimizing the number of nodes in each layer to minimize the overall weight complexity. Consequently, the number of convolution channels will increase, as illustrated in Table I. This optimization can be formulated as:

\begin{equation}
\underset{W}{\text{minimize}} \left( \mathcal{L}(W; \hat{W}, \hat{D}) + \delta \vert W \vert + \delta_g \sum_{e} \vert\vert W^e \vert\vert \right),
\end{equation}
where $ \delta_g $ represents the regularization coefficient and $ e $ denotes a collection of activated connections associated with the newly added nodes. The term $ \delta_g \sum_{e} \vert\vert W^e \vert\vert $ implements filter-wise group LASSO regularization. This technique is crucial for eliminating unnecessary nodes and determining the optimal quantity of nodes across all layers, leading to a more streamlined network.

In conclusion, this process may result in a transient increase in the number of convolutional channels within the network. Consequently, the network is better equipped to extract more pertinent features that are essential for achieving accurate classification results. Through this adaptive approach, our objective is to enhance the overall performance of the model in classifying EEG data.

\begin{table}[!t]
\centering
\renewcommand{\arraystretch}{1}
{
\caption{Detailed results of the test phase. Reported performance is subject-averaged accuracy (\%).}
\resizebox{\columnwidth}{!}{%
\begin{tabular}{c|ccc|c} \hline
                  & Session1 & Session2 & Session3 & Average \\\hline
CSP+RF \cite{qi2012random}          & 23.78    & 22.19    & 26.25    & 23.74   \\
CSP+SVM \cite{FBCSP}         & 25.91    & 29.24    & 32.16    & 29.10   \\
CSP+LDA \cite{FBCSP}         & 17.68    & 23.49    & 22.52    & 21.56   \\
FBCSP \cite{FBCSP}           & 30.00    & 32.17    & 26.74    & 29.30   \\
EEGNet \cite{eegnet}          & 32.13    & 46.88    & 37.63    & 38.88   \\
Deep ConvNet \cite{deepconvnet}    & 39.25    & 49.81    & 45.34    & 44.80   \\
Shallow ConvNet \cite{deepconvnet}  & 45.34    & 46.49    & 43.98    & 45.27   \\
ERA-CNN \cite{ICASSP} & 46.94    & 47.82    & 50.68    & 48.81   \\
MCNN \cite{MCNN}            & 48.53    & 43.45    & 49.73    & 47.90   \\
Proposed method & 50.68    & 56.61    & 57.34    & 54.88  \\
\hline
\end{tabular}}}
\end{table}

\section{Results and Discussion}
The evaluation was conducted in a subject-independent manner, whereby the weights derived from previous sessions contributed to the learning process as each session progressed. Consequently, in Sessions 2 and 3, the weights that demonstrated the highest performance in the previous session were employed, with the initial stages excluded. During this process, the network can be gradually expanded, and the expanded weights are carried over to the subsequent sessions.

Table I summarizes the detailed results of the test phase, presenting the subject-averaged accuracy (\%) across three sessions for each evaluated method. The Proposed method achieved the highest performance, with session accuracies of 50.68 \%, 56.61 \%, and 57.34 \%, resulting in an average of 54.88 \%. This indicates strong adaptability to individual differences in EEG data. ERA-CNN followed with average accuracy of 48.81 \%, showing effective use of temporal information, while Deep ConvNet and Shallow ConvNet achieved average accuracies of 44.80 \% and 45.27 \%, respectively, demonstrating that deeper architectures can capture complex features effectively. In contrast, traditional methods such as CSP+RF (23.74 \%), CSP+SVM (29.10 \%), and CSP+LDA (21.56 \%) showed significantly lower performance. Overall, the proposed method outperforms existing techniques, highlighting the potential of advanced architectures in EEG classification tasks.

To investigate the impact of network expansion on feature extraction within the context of BCIs, t-SNE \cite{van2008visualizing} was employed to visualize the extracted features. According to Fig. 2, when network expansion is applied, the features exhibit a more pronounced clustering compared to scenarios where expansion is not utilized. This enhanced clustering is particularly evident in Session 1 and Session 2 of Fig. 2 (a) and (b), where clear and distinct clusters emerge when expansion is implemented, in contrast to the less organized features observed without it. Fig. 2(c) also demonstrates that applying network expansion results in a more densely clustered feature representation; however, compared to Fig. 2(a) and (b), the degree of clustering is less pronounced.

This observation indicates that the augmented network capacity enables a more effective extraction of pivotal features linked to diverse mental states, which is crucial for precise BCI applications. The capacity to effectively differentiate between categories within the feature space can markedly enhance the performance of classification algorithms utilized in MI classification tasks. In light of the experimental outcomes, this study substantiates that the network expansion strategy not only enhances feature extraction but also holds promise for advancing the efficacy of CNN-based approaches in enhancing the reliability and accuracy of BCI systems.

\section{Conclusion}
The findings of this study provide compelling evidence that an expandable network architecture is an effective method for extracting critical features from EEG signals, particularly in the context of BCIs. The results demonstrate that the proposed method exhibits superior performance compared to traditional techniques, illustrating its capacity to adapt to individual variations in EEG data. Our findings indicate that the network expansion strategy has the potential to mitigate limitations in training performance while enhancing the adaptability and effectiveness of CNN methods in EEG classification tasks. This adaptability is essential for developing robust BCI systems that are tailored to diverse users and their unique neural patterns. Future research should focus on optimizing the expandable architecture to improve its performance across various settings. Additionally, integrating this architecture with real-time BCI systems will be crucial for maximizing user experience and application efficacy. Such advancements will pave the way for innovative BCI solutions that fully leverage EEG signal processing, ultimately enhancing human-technology interaction.
\bibliographystyle{IEEEtran}
\bibliography{Reference}
\end{document}